\documentclass{article}

\usepackage{arxiv}

\usepackage[utf8]{inputenc} % allow utf-8 input
\usepackage[T1]{fontenc}    % use 8-bit T1 fonts
\usepackage{url}            % simple URL typesetting
\usepackage{booktabs}       % professional-quality tables
\usepackage{amsfonts}       % blackboard math symbols
\usepackage{nicefrac}       % compact symbols for 1/2, etc.
\usepackage{microtype}      % microtypography
\usepackage{lipsum}		% Can be removed after putting your text content
\usepackage{graphicx}
\usepackage{subcaption}
\usepackage{booktabs}
\usepackage{mwe}
\usepackage{listings}
\usepackage{caption}
\usepackage{amsmath}
\usepackage{longtable}
\usepackage{todonotes}

\DeclareFixedFont{\ttb}{T1}{txtt}{bx}{n}{12} % for bold
\DeclareFixedFont{\ttm}{T1}{txtt}{m}{n}{12}  % for normal

\usepackage{hyperref}

\title{Towards a Network Control Theory of Electroconvulsive Therapy Response}

\date{}

\author{
Tim Hahn\textsuperscript{1}\thanks{
	These authors contributed equally	
}
\And Hamidreza Jamalabadi\textsuperscript{2$\ast$}
\And Erfan Nozari\textsuperscript{3}
\And Nils R. Winter\textsuperscript{1}
\And Jan Ernsting\textsuperscript{1,4}
\And Marius Gruber\textsuperscript{1}
\And Marco J. Mauritz\textsuperscript{1}
\And Pascal Grumbach\textsuperscript{1}
\And Lukas Fisch\textsuperscript{1}
\And Ramona Leenings\textsuperscript{1,4}
\And Kelvin Sarink\textsuperscript{1}
\And Julian Blanke\textsuperscript{1}
\And Leon Kleine Vennekate\textsuperscript{1}
\And Daniel Emden\textsuperscript{1}
\And Nils Opel\textsuperscript{1}
\And Dominik Grotegerd\textsuperscript{1}
\And Verena Enneking\textsuperscript{1}
\And Susanne Meinert\textsuperscript{1,5}
\And Tiana Borgers\textsuperscript{1}
\And Melissa Klug\textsuperscript{1}
\And Elisabeth J. Leehr\textsuperscript{1}
\And Katharina Dohm\textsuperscript{1}
\And Walter Heindel\textsuperscript{6}
\And Joachim Gross\textsuperscript{7}
\And Udo Dannlowski\textsuperscript{1}
\And Ronny Redlich\textsuperscript{1,8$\ast$}
\And Jonathan Repple\textsuperscript{1$\ast$}
}

\begin{document}
\maketitle

\section*{Affiliations}
	1 Institute for Translational Psychiatry, University of Münster, Germany
\newline
2 Department of Psychiatry and Psychotherapy, University of Tübingen, Germany
\newline
3 Department of Mechanical Engineering, University of California, Riverside, USA
\newline
4 Faculty of Mathematics and Computer Science, University of Münster, Germany 
\newline
5 Institute for Translational Neuroscience, University of Münster, Germany 
\newline
6 Institute of Clinical Radiology, University of Münster, Germany
\newline
7 Institute for Biomagnetism and Biosignalanalysis, University Hospital Münster, Germany
\newline
8 Department of Psychology, University of Halle, Germany
\pagebreak

\begin{abstract}
Electroconvulsive Therapy (ECT) is arguably the most effective intervention for treatment-resistant depression. While large interindividual variability exists, a theory capable of predicting individual response to ECT remains elusive. To address this, we posit a quantitative, mechanistic framework of ECT response based on Network Control Theory (NCT). Then, we empirically test our approach and employ it to predict ECT treatment response. To this end, we derive a formal association between Postictal Suppression Index (PSI) – an ECT seizure quality index – and whole-brain modal and average controllability, NCT metrics based on white matter brain network architecture, respectively. Exploiting the known association of ECT response and PSI, we then hypothesized an association between our controllability metrics and ECT response mediated by PSI. We formally tested this conjecture in N=50 depressive patients undergoing ECT. We show that whole-brain controllability metrics based on pre-ECT structural connectome data predict ECT response in accordance with our hypotheses. In addition, we show the expected mediation effects via PSI. Importantly, our theoretically motivated metrics are at least on par with extensive machine learning models based on pre-ECT connectome data. In summary, we derived and tested a control-theoretic framework capable of predicting ECT response based on individual brain network architecture. It makes testable, quantitative predictions regarding individual therapeutic response, which are corroborated by strong empirical evidence. Our work might constitute a starting point for a comprehensive, quantitative theory of personalized ECT interventions rooted in control theory.
\end{abstract}

% keywords can be removed
% \keywords{First keyword \and Second keyword \and More}

\section{Introduction}
Electroconvulsive therapy (ECT) is the most effective treatment for severe and therapy-resistant depression.[1–3] As therapeutic response varies widely, however, numerous studies have sought to predict individual ECT response from indexes, rating scales, or symptom clusters prior to intervention, among others.[4–7] Although limited to very small patient samples (N=24), more recent machine learning models based on neuroimaging data have rekindled hopes for a theranostic biomarker.[8] Despite these efforts, a theoretical framework capable of predicting ECT response on an individual level remains elusive.

Generally, during ECT an electric charge is applied to the brain to induce a generalized tonic-clonic seizure characterized by high amplitudes and polyspike waves as well as high amplitudes with slow waves in the electroencephalogram (EEG). This is followed by the termination phase during which postictal suppression – i.e., a period of lower spectral amplitude and a general flattening of the EEG – occurs.[9] Empirically, postictal suppression has repeatedly been linked to treatment response: A plethora of studies for more than three decades has found a positive association between stronger postictal suppression and better response to ECT.[9, 10, 19, 20, 11–18] Recognizing this, ECT devices today provide a measure of postictal suppression which is routinely used in clinical practice to assess treatment quality and estimate future therapeutic response. This postictal suppression index (PSI) is quantified by the ratio of EEG signal power during the termination phase and the tonic-clonic seizure and subtracting it from 1 (see equation 1 in Methods).

While empirically well-founded and routinely used in the clinic, a theoretical framework explaining why ECT patients vary with regard to seizure quality indices such as PSI is, however, missing. Consequently, PSI is of limited theranostic utility as it can only be measured during ECT – i.e., after the beginning of the intervention – and thus cannot serve as a predictive marker in therapeutic planning. In contrast, identifying the inter-individual differences underlying the observed variance in PSI would allow us to derive a predictive marker for ECT response suitable for therapeutic planning.

To enable the theoretically driven construction of a theranostic biomarker, we conceptualize ECT as an attempt to drive the brain into a specific state (i.e., seizure) by influencing large-scale, dynamic network state transitions in the brain. Building on Control Theory as the study and practice of controlling dynamical systems[21], we can then view the electric charge applied during ECT as a control input designed to guide the system towards a specific state – i.e., a seizure – characterized by the high amplitude EEG oscillations typical for tonic-clonic seizures. 

Next, we aim to relate this control theoretic perspective to network neuroscience: Specifically, recent progress in Network Control Theory has enabled the quantification of the influence a brain region (or “node” from a network perspective) has on the dynamic transitions between brain states. [22, 23] This so-called controllability of a brain region is linked to the system’s structural connectivity properties (usually derived from Diffusion Tensor Imaging (DTI) data) which constrain or support transitions between different brain states [24, 25]. Controllability is commonly captured by two key metrics: On the one hand, modal controllability represents the ability to control especially fast decaying neural dynamics.[26] On the other hand, average controllability measures the ability of a system to spread and amplify the control inputs (i.e., the electric charge used during ECT) and is thus indicative of a node’s ability to support low-energy state transitions. For formal definitions of average and modal controllability measures see Methods section. Fueled by evidence that the human brain is in principle controllable [24] and the recently discovered associations with cognition [25, 27], numerous studies have empirically investigated the two metrics in mental disorders.[28–30]  Focusing specifically on Major Depressive Disorder (MDD), Hahn et al. recently showed in a large sample of patients that whole-brain average and modal controllability (i.e., mean average and modal controllability across the brain) is related to genetic, individual, and familial risk in MDD patients.[31] 

In the current study we formally derive from control theory that whole-brain modal and average controllability are mathematically related to amplitude of global brain response to control input (i.e., output signal power after a control input): As whole-brain average controllability is defined as the weighted sum of the so-called impulse response of each brain area, it is directly related to the cumulative power of the output response (i.e., the EEG signal power during the seizure). Likewise, whole-brain modal controllability is linked to the observed brain dynamics after inducing energy into the system. For a formal derivation of these properties see Supplementary Methods section.

In other words, if we conceptualize the electric charge applied during ECT as a control input designed to induce a seizure, it follows that whole-brain average and modal controllability derived from the structural connectome are expected to be related to EEG signal power after control input (i.e., during the induced seizure). Specifically, higher whole-brain average and lower modal controllability ought to be related to higher output signal power during the tonic-clonic seizure (see Network Control Analysis in the Methods section for a more stringent argument). Importantly, as PSI is proportional to the ratio of signal power during the termination phase and the tonic-clonic seizure (see equation 1 in Methods), higher signal power during the tonic-clonic seizure phase should result in higher PSI.

In summary, we posit that lower whole-brain modal controllability and higher whole-brain average controllability (both mathematically associated with higher output signal power during the tonic-clonic seizure as formally derived using the principles of Control Theory) should lead to higher PSI values. As the positive relationship between PSI and therapeutic response is well-documented, it follows that lower whole-brain modal controllability and higher whole-brain average controllability, respectively, should result in higher PSI and thus improved ECT response. Here, we empirically test this conjecture by assessing the relation between empirically observed PSI during ECT and whole-brain modal and average controllability derived from pre-ECT structural connectome data. Next, we replicate the association between PSI and ECT response known from the literature. Then, we test whether pre-ECT whole-brain modal and average controllability predict ECT response and assess whether these effects are indeed mediated by PSI during ECT. Finally, we compare the predictive power of whole-brain modal and average controllability regarding therapeutic response to the performance of an extensive array of machine learning models based on structural connectome data.

\section{Methods}
\subsection{Sample}
Fifty-three participants diagnosed with current major depressive disorder and treated with electroconvulsive therapy participated in the present study. Patients were recruited naturalistically with treatment being assigned based on clinical decisions independent from study participation. All subjects were diagnosed with the Structured Clinical Interview for DSM-IV-TR [32] to confirm the psychiatric diagnosis. For medication details and study inclusion and exclusion criteria see Supplements. In the process of image quality control (see methods section below), three subjects had to be excluded. Therefore, the final sample comprised of 50 subjects (29 female, 21 male; mean age= 45.1 years, SD= 10.8). Note that results do not substantially change if these three subjects are not excluded. This study was approved by the ethics committee of the Medical Faculty of Muenster University and all subjects gave written informed consent prior to participation. They received financial compensation after study completion.

For MRI data acquisition and preprocessing, see Supplementary Methods.

\subsection{Electroconvulsive Therapy}
Brief-pulse ECT was conducted three times a week using an integrated instrument (Thymatron IV; Somatics Inc; number of sessions: M=13.0, SD=4.34, range=5-25). Energy dosage elevation was considered between ECT sessions if the primarily induced seizure activity lasted less than 25 seconds. For more details on ECT procedure and parameters see Supplements. 

Recognizing the empirical link between PSI and ECT response, many ECT devices today provide a measure of postictal suppression which is routinely used in clinical practice to assess seizure quality and estimate future therapeutic response. During ECT, the ictal outcome of PSI was measured and summarized by the Thymatron machines using five electrodes. These were placed on the right and left forehead, on the right and left mastoid, and on on the patient's nasion (ground electrode). The PSI is quantified by the ratio of signal power during the termination phase and power during the tonic-clonic seizure based on the EEG time-series and subtracting it from 1: 
\begin{align}
	%PSI=(1-(Power_(termination phase))/(Power_(tonic-clonic seizure) ))∝Power_(tonic-clonic seizure)  				(1)
	PSI = \left(1-\frac{\text{Power}_\text{termination phase}}{\text{Power}_\text{tonic-clonic seizure}}\right) \propto \text{Power}_\text{tonic-clonic seizure}
\end{align}
It follows that higher EEG signal power during the tonic-clonic seizure should result in higher PSI values.
 
We used Thymatron system IV’s default settings to compute PSI. Specifically, signal power during the termination phase and the seizure, respectively, is computed as the mean power of three 1.28 second long intervals with an EEG sampling rate of 200 Hz. To minimize artifacts, 3.84 seconds around the tonic-clonic seizure endpoint are disregarded (F. Berninger, Somatics Inc. representative, personal e-mail communication with JR, December 15, 2017).

\subsection{Choice of primary measure}
In this work, we used ECT response as the primary outcome measure. MDD symptoms at both time points were measured using the Hamilton Depression Rating Scale (HDRS).[33]  ECT response was defined as the difference in MDD symptoms before and after ECT. Specifically, we subtracted pre-ECT scores from post-ECT scores so that improvement is indicated by more negative values. Despite concerns regarding its psychometric properties[34], the HDRS is one of the most widely used clinician-administered depression assessment scale and is routinely used in studies investigating ECT response.[8, 35] 

\subsection{Statistical and Machine Learning Analyses}
To empirically test our hypotheses, we conducted four analyses: First, we assessed the relation between empirically observed whole-brain modal $(\overline{MC})$ and average controllability $(\overline{AC})$, respectively, and PSI during ECT. To this end, we employed an ANCOVA approach with PSI as the dependent variable, $\overline{MC}$ and $\overline{AC}$, respectively, as the independent variable. Second, to replicate the association between PSI and ECT response known from the literature, we employed an ANCOVA approach with ECT response as the dependent variable and PSI as the independent variable. Third, we tested whether pre-ECT $\overline{MC}$ and $\overline{AC}$ predicted ECT response. Again, we employed an ANCOVA approach with ECT response as the dependent variable and $\overline{MC}$ and $\overline{AC}$, respectively, as the independent variable. Note that in all of these analyses, in addition to the F-value and the one-sided p-value, we provide $\eta p^2$ as a measure of effect size. All of these analyses were conducted using the Python statsmodels package (statsmodels.org).

Fourth, we tested if this effect of $\overline{MC}$ and $\overline{AC}$, respectively, on ECT response is indeed mediated by PSI during ECT. To this end, we employed a mediation analysis with $\overline{MC}$ and $\overline{AC}$ as the predictor, PSI as the mediator, and ECT response as the outcome using the bias-correct, non-parametric, bootstrap-based mediation analysis implemented in the Python pingouin library (pingouin-stats.org). Significance of the indirect (mediation) effect was computed using the permutation test as outlined in Mac Kinnon[36] (Section 12.6) based on 10,000 permutations. 

In all analyses, we included age, sex, and symptom severity before ECT as covariates. To ensure that the effects observed are not driven by basic graph properties, we additionally included the number of present edges in all analyses. In analyses involving PSI as computed from EEG during ECT, we only used data from the first ECT session as numerous studies have shown that the structural connectome as measured with DTI underlying our control analyses changes in response to previous ECT sessions.[37] Note that PSI was available for 45 of the 50 patients. Thus, all analyses involving PSI are based on N=45, whereas all other analyses are based on N=50. 

Finally, we conducted an extensive array of machine learning analyses based on structural connectome data – namely Fractional Anisotropy, Mean Diffusivity, and Number of Streamlines – to predict ECT response. Using the PHOTON AI software[38], we trained and evaluated a total of 35 machine learning pipelines. Specifically, we tested pipelines including Principal Component Analysis and univariate feature selection (with 5\% and 10\% thresholds) with linear and non-linear Support Vector Machines, Random Forests as well as Linear Regression. Models were trained and evaluated in a nested leave-one-out cross-validation framework and performance reported as percent variance explained by Spearman rank correlation between true and predicted treatment response. The full code can be found in the Supplementary Material. We compared this to a simple linear regression model also using leave-one-out cross-validation based solely on the single $\overline{MC}$ and $\overline{AC}$ value, respectively, without any further optimization or model selection. Note that formally testing performance differences for significance using e.g. 1000 permutations would require fitting 257,750,000 machine learning pipelines for the three structural connectome modalities alone and was thus not feasible given current hardware. 

\section{Results}
To empirically test our hypotheses derived via the application of control theory to ECT, we conducted four analyses: First, we assessed the effect of the empirically observed whole-brain modal controllability $(\overline{MC})$ and whole-brain average controllability $(\overline{AC})$, respectively, on PSI during ECT. As hypothesized, $\overline{MC}$  is negatively associated with PSI $(F(1,39)=3.28, p=0.039, \text{partial } \eta^2 (\eta p^2)=0.077)$. Likewise, $\overline{AC}$ shows a trend-wise positive association with PSI $(F(1,39)=1.72, p=0.098, \eta p^2=0.042)$. 

Second, we replicated the association between higher PSI and better ECT response known from the literature $(F(1,39)=7.28, p=0.005, \eta p^2=0.157)$.

Third, we tested whether pre-ECT $\overline{MC}$ and $\overline{AC}$ predicted ECT response. In line with our hypotheses, we reveal that lower $\overline{MC}$ is indeed associated with stronger ECT response $(F(1,44)=8.15, p=0.004, \eta p^2=0.156)$. Furthermore, as expected, $\overline{AC}$ is positively related to ECT response $(F(1,44)=3.62, p=0.032, \eta p^2=0.076)$. 

\begin{figure}
	\includegraphics{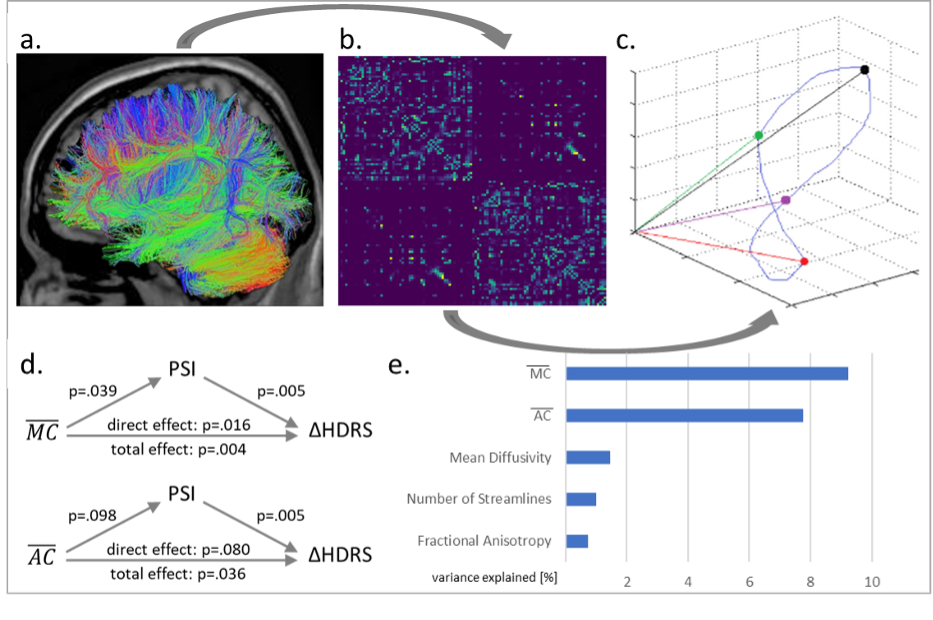}
	\caption{From Diffusion Tensor Imaging (DTI) data (a), we derived the structural connectivity matrix for each patient (b) and quantified modal and average controllability – i.e., the influence a brain region has on the dynamic transitions between brain states underlying cognition and behavior (illustrated in c). We then show that mean modal ($\overline{MC})$ and average controllability ($\overline{AC}$), respectively, are associated with therapeutic response to ECT treatment and that this effect is mediated by Postictal Suppression Index (PSI; d). $\overline{MC}$ and $\overline{AC}$ also predict therapeutic response in patients before ECT treatment as good or better than an extensive array of multivariate machine learning models based on Fractional Anisotropy, Mean Diffusivity, and Number of Streamlines derived from DTI (e).}
	\label{fig1}
\end{figure}

Fourth, we tested if the effects of $\overline{MC}$ and $\overline{AC}$, respectively, on ECT response are indeed mediated by PSI. To this end, we employed a mediation analysis with $\overline{MC}$ and $\overline{AC}$, respectively, as the predictor, PSI as the mediator, and ECT response as the outcome (Figure \ref{fig1}d). We indeed observed a significant mediation (i.e., indirect) effect for $\overline{MC}$ $(ab=184.80; p=0.010)$ as well as for $\overline{AC}$ $(ab=-47.94; p=0.017)$. However, a significant direct effect (c’) – i.e., an association between $\overline{MC}$ with ECT response after controlling for PSI – remains $(c’=706.26; p=0.016)$; indicating a relation to ECT response above and beyond the PSI-mediated effect. In contrast, the effect of $\overline{AC}$ on ECT response is fully mediated by PSI, i.e., it no longer reaches significance when controlling for PSI $(c’= -140.40; p=0.080)$.

Finally, we descriptively compared the predictive power of $\overline{MC}$ and $\overline{AC}$ regarding ECT response to the performance of an extensive array of machine learning models based on structural connectome data. Specifically, we tested 35 combinations of data transformations (including Principal Component Analysis and univariate feature selection) and multivariate machine learning algorithms (including linear and non-linear Support Vector Machines, Random Forests, and linear Regression Models) in a nested cross-validation approach using Fractional Anisotropy, Mean Diffusivity, and Number of Streamlines derived from each patient’s structural connectome, respectively (see Methods for details). We showed that  $\overline{MC}$ $(r^2=9.24\%)$ as well as $\overline{AC}$ $(r^2=7.76\%)$, respectively, explain nominally more variance in treatment response than the best machine learning model for each modality (Fractional Anisotropy $r^2=0.74\%$; Mean Diffusivity $r^2=1.46\%$; Number of Streamlines $r^2=0.99\%$; Figure \ref{fig1}e).

\section{Discussion}
In this work, we drew upon Network Control Theory to derive a mechanistic framework capable of predicting ECT response based on individual, pre-ECT white matter architecture. This approach has two main strengths: First, unlike previous efforts in psychiatry, it allows us to derive testable predictions regarding individual therapeutic response from a well-established, quantitative theory. Second, these theoretically derived predictions – which, unlike PSI, can be obtained before ECT and can thus be used for treatment planning – are at least on par with state-of-the-art multivariate machine learning approaches.

Specifically, we first derive that lower whole-brain modal controllability and higher whole-brain average controllability are mathematically associated with higher output signal power during the tonic-clonic seizure if we conceptualize the electric charge applied during ECT as the control input to a noise-free, time-invariant linear system commonly used in NCT and the seizure as its output. Based on the calculation of PSI as routinely done in the clinical context, we can then directly deduce that – all other things being equal – PSI should increase for patients with lower $\overline{MC}$ and higher $\overline{AC}$, respectively. Drawing on previous evidence showing a positive association between PSI and ECT response, we then hypothesize higher ECT response for patients with lower $\overline{MC}$ and higher $\overline{AC}$, respectively. All three effects predicted by our approach for $\overline{MC}$ – i.e., the associations between 1. $\overline{MC}$ and PSI, 2. PSI and ECT response, and 3. $\overline{MC}$ and ECT response – as well as the mediation effect of PSI, were empirically found in the theoretically expected directions. For $\overline{AC}$, results were also in the expected direction, but effects were smaller ($\eta p^2 of 0.068 and 0.076$) and in case of the association of $\overline{AC}$ and PSI only marginally significant. These findings support the notion that ECT response – at least in part – can be understood analogous to physical dynamical systems within the formal framework of Network Control Theory viewing the electric charge applied during ECT as the control input and the ECT-induced seizure as the control output. 

Importantly, a comparison with state-of-the-art machine learning approaches showed that a simple linear predictive model using only the single $\overline{AC}$ or $\overline{MC}$ value, respectively, for each patient is at least as good a predictor of therapeutic response as an extensive array of 35 machine learning pipelines based on the multivariate patterns of Fractional Anisotropy, Mean Diffusivity, and Number of Streamlines derived from each patient’s structural connectome. Generally, this shows that our theoretically derived metrics are comparable or better predictors of treatment response than an extensive array of state-of-the-art multivariate machine learning approaches – thus encouraging further research for theranostic markers in this direction.

More generally, connecting Network Control Theory and ECT opens the door towards a quantitative understanding of ECT response. Providing a quantitative answer to the question of what changes in the brain after a specified stimulation event is not only crucial for ECT response prediction, but also – with Network Control Theory – provides a rich theoretical framework with which we can hope to optimize ECT application itself. For example, our results are conceptually related to successful attempts to predict stimulation outcome in the context of electrical brain stimulation.[39] While demonstrating that variation in response to treatment can be explained by controllability differences, our approach, however, relies exclusively on whole-brain controllability metrics – not local controllability of certain brain regions. While we think this is particularly reasonable in the context of ECT, which induces a generalized seizure across the entire brain, future studies ought to extend the idea to include localized control. This is particularly interesting as Network Control Theory provides a straightforward way to calculate which brain regions should be targeted in which order to efficiently reach a specific state (be it a seizure or other states) employing more localized interventions such as Transcranial Magnetic Stimulation. Whether targeting regions identified by this so-called control-node analysis[40] leads to stronger response can be directly tested in future intervention studies. Importantly, this view also entails that therapeutic interventions should be custom-tailored to the patient’s individual structural connectome topology which might thus serve as a promising approach to patient stratification.

In addition to the prediction of ECT response, a control-theoretic framework might help to understand several phenomena associated with ECT. For example, inducing seizures becomes increasingly difficult with age which might be explained by the fact that synchronizability – i.e., the ability of the constituents of a dynamical system to show coherent activity – develops in a way as to favor seizure suppression over the lifespan.[41] Also, the same study shows that strong modal controllers were disproportionately located in cognitive control systems, including both the frontoparietal and cingulo-opercular systems. This might entail that the efficacy of ECT and its adverse effects regarding cognitive deficits maybe linked at the level of white-matter network architecture. 

Several limitations should be noted. First, calculation of average and modal controllability relies on the simplified noise-free linear discrete-time and time-invariant network model employed in virtually all work on brain Network Control Theory [24, 42, 43]. Given the brain’s clearly non-linear dynamics, this is justified as 1) nonlinear behavior may be accurately approximated by linear behavior [44, 45] and 2) the controllability of linear and nonlinear systems is related such that a controllable linearized system is locally controllable in the nonlinear case (see also [24] for details).

Second, our estimation of controllability is based upon DTI tractography which itself is limited in its ability to accurately quantify the structural connectome (for an introduction, see [46]). Currently, several novel approaches to controllability quantification are being explored including estimation from gray matter [47] and resting-state functional dynamics [42]. Empirically comparing and theoretically reconciling results from these methods will be crucial for robust parameter estimation in Network Control Theory studies of the brain. In addition, longitudinal data from DTI, gray matter, and resting-state functional dynamics available from e.g. the Marburg-Münster Affective Disorders Cohort Study (MACS; [48]) will enable us to assess the (differential) reliability of these approaches. In combination with functional Magnetic Resonance Imaging (fMRI), this also provides an opportunity to further characterize the relationship between network control and individual task-related activation [49].

In summary, we derived and tested a control-theoretic framework capable of predicting ECT response based on individual brain network architecture. It makes testable, quantitative predictions regarding individual therapeutic response, which are corroborated by strong empirical evidence. Our work might constitute a starting point for a comprehensive, quantitative theory of personalized ECT interventions rooted in control theory.

\section*{Funding}
This work was funded by the German Research Foundation (DFG grants RE4458/1-1 to RR, HA7070/2-2, HA7070/3, HA7070/4 to TH) and the Interdisciplinary Center for Clinical Research (IZKF) of the medical faculty of Münster (grants Dan3/012/17 to UD , MzH 3/020/20 to TH and SEED 11/19 to NO) as well as the “Innovative Medizinische Forschung” (IMF) of the medical faculty of Münster (Grants OP121710 to NO and TH, RE111722 to RR). HJ was supported by Fortüne grant of Medical Faculty of University of Tübingen (No. 2487-1- 0). The study sponsors had not role in the study design, the collection, analysis, and interpretation of data, writing of the report or in the decision to submit the paper for publication.

\section*{References}
 1. 	Kellner CH, Greenberg RM, Murrough JW, Bryson EO, Briggs MC, Pasculli RM. ECT in treatment-resistant depression. Am J Psychiatry. 2012;169:1238–1244.
 
2. 	Van Den Broek WW, De Lely A, Mulder PGH, Birkenhäger TK, Bruijn JA. Effect of antidepressant medication resistance on short-term response to electroconvulsive therapy. J Clin Psychopharmacol. 2004;24:400–403.

3. 	Geddes J, Carney S, Cowen P, Goodwin G, Rogers R, Dearness K, et al. Efficacy and safety of electroconvulsive therapy in depressive disorders: A systematic review and meta-analysis. Lancet. 2003;361:799–808.

4. 	De Vreede IM, Burger H, Van Vliet IM. Prediction of response to ECT with routinely collected data in major depression. J Affect Disord. 2005;86:323–327.

5. 	Abrams R, Vedak C. Prediction of ECT Response in Melancholia. Convuls Ther. 1991;7:81–84.

6. 	Hickie I, Parsonage B, Parker G. Prediction of response to electroconvulsive therapy. Preliminary validation of a sign-based typology of depression. Br J Psychiatry. 1990;157:65–71.

7. 	Hickie I, Mason C, Parker G, Brodaty H. Prediction of ECT response: Validation of a refined sign-based (CORE) system for defining melancholia. Br J Psychiatry. 1996;169:68–74.

8. 	Redlich R, Opel N, Grotegerd D, Dohm K, Zaremba D, Burger C, et al. Prediction of individual response to electroconvulsive therapy via machine learning on structural magnetic resonance imaging data. JAMA Psychiatry. 2016;73:557–564.

9. 	Perera TD, Luber B, Nobler MS, Prudic J, Anderson C, Sackeim HA. Seizure expression during electroconvulsive therapy: Relationships with clinical outcome and cognitive side effects. Neuropsychopharmacology. 2004;29:813–825.

10. 	Luber B, Nobler MS, Moeller JR, Katzman GP, Prudic J, Devanand DP, et al. Quantitative EEG during seizures induced by electroconvulsive therapy: Relations to treatment modality and clinical features. II. Topographic analyses. J ECT. 2000;16:229–243.

11. 	Nobler MS, Sackeim HA. Neurobiological correlates of the cognitive side effects of electroconvulsive therapy. J ECT. 2008;24.

12. 	Nobler MS, Luber B, Moeller JR, Katzman GP, Prudic J, Devanand DP, et al. Quantitative EEG during seizures induced by electroconvulsive therapy: Relations to treatment modality and clinical features. I. Global analyses. J ECT. 2000;16.

13. 	Krystal AD, Weiner RD, Coffey CE, Smith P, Arias R, Moffett E. EEG evidence of more ‘intense’ seizure activity with bilateral ECT. Biol Psychiatry. 1992;31.

14. 	Krystal AD, Weiner RD, Coffey CE. The ictal EEG as a marker of adequate stimulus intensity with unilateral ECT. J Neuropsychiatry Clin Neurosci. 1995;7.

15. 	Krystal AD, Weiner RD. ECT seizure therapeutic adequacy. Convuls Ther. 1994;10.

16. 	Krystal AD, Weiner RD, Swartz CM. Low-frequency ictal EEG activity and ECT therapeutic impact. Convuls Ther. 1993;9.

17. 	Krystal AD. The clinical utility of ictal EEG seizure adequacy models. Psychiatr Ann. 1998;28.

18. 	Sackeim HA, Prudic J, Devanand DP, Nobler MS, Lisanby SH, Peyser S, et al. A prospective, randomized, double-blind comparison of bilateral and right unilateral electroconvulsive therapy at different stimulus intensities. Arch Gen Psychiatry. 2000;57.

19. 	Sackeim HA, Luber B, Katzman GP, Moeller JR, Prudic J, Devanand DP, et al. The effects of electroconvulsive therapy on quantitative electroencephalograms: Relationship to clinical outcome. Arch Gen Psychiatry. 1996;53.

20. 	Nobler MS, Sackeim HA, Solomou M, Luber B, Devanand DP, Prudic J. EEG manifestations during ECT: effects of electrode placement and stimulus intensity. Biol Psychiatry. 1993;34.

21. 	Thomas PJ, Olufsen M, Sepulchre R, Iglesias PA, Ijspeert A, Srinivasan M. Control theory in biology and medicine: Introduction to the special issue. Biol Cybern. 2019;113:1–6.

22. 	Kim JZ, Bassett DS. Linear Dynamics \& Control of Brain Networks. 2019:1–30.

23. 	Braun U, Schaefer A, Betzel RF, Tost H, Meyer-Lindenberg A, Bassett DS. From Maps to Multi-dimensional Network Mechanisms of Mental Disorders. Neuron. 2018;97:14–31.

24. 	Gu S, Pasqualetti F, Cieslak M, Telesford QK, Yu AB, Kahn AE, et al. Controllability of structural brain networks. Nat Commun. 2015;6:1–10.

25. 	Lee WH, Rodrigue A, Glahn DC, Bassett DS, Frangou S. Heritability and Cognitive Relevance of Structural Brain Controllability. Cereb Cortex. 2020;30:3044–3054.

26. 	Karrer TM, Kim JZ, Stiso J, Kahn AE, Pasqualetti F, Habel U, et al. A practical guide to methodological considerations in the controllability. J Neural Eng. 2020. 2020.

27. 	Cornblath EJ, Tang E, Baum GL, Moore TM, Adebimpe A, Roalf DR, et al. Sex differences in network controllability as a predictor of executive function in youth. Neuroimage. 2019;188:122–134.

28. 	Jeganathan J, Perry A, Bassett DS, Roberts G, Mitchell PB, Breakspear M. Fronto-limbic dysconnectivity leads to impaired brain network controllability in young people with bipolar disorder and those at high genetic risk. NeuroImage Clin. 2018;19:71–81.

29. 	Braun U, Harneit A, Pergola G, Menara T, Schaefer A, Betzel RF, et al. Brain state stability during working memory is explained by network control theory, modulated by dopamine D1/D2 receptor function, and diminished in schizophrenia. 2019. 21 June 2019.

30. 	Parkes L, Moore TM, Calkins ME, Cieslak M, Roalf DR, Wolf DH, et al. Network controllability in transmodal cortex predicts psychosis spectrum symptoms. 2020. 1 October 2020.

31. 	Hahn T, Winter NR, Ernsting J, Gruber M, Mauritz MJ, Fisch L, et al. Genetic, Individual, and Familial Risk Correlates of Brain Network Controllability in Major Depressive Disorder.

32. 	Wittchen H, Zaudig M, Fydrich T. Strukturiertes Klinisches Interview für DSM-IV, Hogrefe. 1997.

33. 	HAMILTON M. A rating scale for depression. J Neurol Neurosurg Psychiatry. 1960;23:56–62.

34. 	Michael Bagby R, Ryder AG, Deborah Schuller MR, Marshall MB. The Hamilton Depression Rating Scale: Has the Gold Standard Become a Lead Weight? vol. 161. 2004.

35. 	Repple J, Meinert S, Bollettini I, Grotegerd D, Redlich R, Zaremba D, et al. Influence of electroconvulsive therapy on white matter structure in a diffusion tensor imaging study. Psychol Med. 2020;50:849–856.

36. 	MacKinnon DP. Introduction to statistical mediation analysis. 1st Editio. Taylor \& Francis Inc; 2008.

37. 	J Z, Q L, L D, W L, Y L, H L, et al. Reorganization of Anatomical Connectome following Electroconvulsive Therapy in Major Depressive Disorder. Neural Plast. 2015;2015.

38. 	Leenings R, Winter NR, Plagwitz L, Holstein V, Ernsting J, Steenweg J, et al. Photon-A python api for rapid machine learning model development. ArXiv. 2020.

39. 	Stiso J, Khambhati AN, Menara T, Kahn AE, Stein JM, Das SR, et al. White Matter Network Architecture Guides Direct Electrical Stimulation through Optimal State Transitions. Cell Rep. 2019;28:2554-2566.e7.

40. 	Brunton SL, Kutz JN. Data-Driven Science and Engineering. Cambridge University Press; 2019.

41. 	Tang E, Giusti C, Baum GL, Gu S, Pollock E, Kahn AE, et al. Developmental increases in white matter network controllability support a growing diversity of brain dynamics. Nat Commun. 2017;8:1252.

42. 	Gu S, Deng S. Controllability Analysis on Functional Brain Networks. 2019:1–26.

43. 	Lynn CW, Bassett DS. The physics of brain network structure, function and control. Nat Rev Phys. 2019;1:318–332.

44. 	Honey CJ, Sporns O, Cammoun L, Gigandet X, Thiran JP, Meuli R, et al. Predicting human resting-state functional connectivity from structural connectivity. Proc Natl Acad Sci U S A. 2009;106:2035–2040.

45. 	Nozari E, Bertolero MA, Stiso J, Caciagli L, Cornblath EJ, He X, et al. Is the brain macroscopically linear? A system identification of resting state dynamics. 2020. 22 December 2020.

46. 	Jbabdi S, Johansen-Berg H. Tractography: Where Do We Go from Here? Brain Connect. 2011;1:169–183.

47. 	Jamalabadi H, Zuberer A, Kumar VJ, Li M, Alizadeh S, Moradi AA, et al. The missing role of gray matter in studying brain controllability. https://doi.org/10.1101/2020.04.07.030015.

48. 	Vogelbacher C, Möbius TWD, Sommer J, Schuster V, Dannlowski U, Kircher T, et al. The Marburg-Münster Affective Disorders Cohort Study (MACS): A quality assurance protocol for MR neuroimaging data. Neuroimage. 2018;172:450–460.

49. 	Taghia J, Cai W, Ryali S, Kochalka J, Nicholas J, Chen T, et al. Uncovering hidden brain state dynamics that regulate performance and decision-making during cognition. Nat Commun. 2018;9:1–19.

\newpage
\section{Supplementary Material}

\subsection{Supplementary Methods}

\subsubsection{MRI data acquisition}
Data was acquired before the first ECT session using a 3T whole body MRI scanner (Gyroscan Intera, Philips Medical Systems, Best, the Netherlands), as reported earlier [1]. 

T1-weighted high resolution anatomical images were acquired with a 3D fast gradient echo sequence ('Turbo Field Echo', TFE), TR=7.4 ms, TE=3.4 ms, FA=9°, 2 signal averages, inversion prepulse every 814.5ms, acquired over a field of view of 256(FH)x204(AP)x160(RL) mm, phase encoding in AP and RL direction, reconstructed to cubic voxels of 0.5x0.5x0.5 mm.

The DTI data was acquired in 36 axial slices, 3.6 mm thick with no gap (acquired matrix 128 x 128), resulting in a voxel size of $1.8 x 1.8 x 3.6 mm^3$. The echo time was 95 ms and the repetition time was 9473 ms. A b-value of $1000 sec/mm^2$ was used for 20 diffusion-weighted images, with isotropic gradient directions plus one non-diffusion-weighted ($b = 0 s/mm^2$) image. In sum, 21 images per slice were used for diffusion-tensor estimation. The total data acquisition time was approximately 8 minutes per subject. During the experiment, subjects lay supine in the MRI scanner with their head position being stabilized (Figure \ref{fig1}a).

\subsubsection{Imaging Data Preprocessing}
Connectomes were reconstructed using the CATO toolbox[2] following the procedure outlined in [3]. For a more detailed description of the preprocessing see [4]. In accordance with [4], we decided on using a basic DTI reconstruction rather than more advanced diffusion direction reconstruction methods to provide a reasonable balance between false negative and false positive fiber reconstructions [5]. For each subject an anatomical brain network was reconstructed, consisting of 114 cortical areas of a subdivision of the FreeSurfer’s Desikan–Killiany atlas [6, 7], and the reconstructed streamlines between these areas  (Figure 1b). White matter connections were reconstructed using deterministic streamline tractography, based on the Fiber Assignment by Continuous Tracking (FACT) algorithm [8]. Network connections were included when two nodes (i.e., brain regions) were connected by at least three tractography streamlines [9]. Note that results do not substantially change if this threshold is omitted or if this threshold is increased to 10 connections.  Also note that including head motion parameters from DTI did not substantially change the results. For each participant, the network information was stored in a structural connectivity matrix, with rows and columns reflecting cortical brain regions, and matrix entries representing graph edges. Edges were only described by their presence or absence to create unweighted graphs and scaled to ensure LTI model stability (for DTI quality control, see Supplementary Material).

\subsubsection{Network Control Analysis}
Building on Control Theory as the study and practice of controlling dynamical systems[10], we view the electric charge applied during ECT as the control input (u) designed to guide the system towards a seizure which characterize the output. Against this background, we apply the standard model of structural brain controllability assuming a noise-free linear time-invariant model (for a more detailed introduction, see [11, 12]):
\begin{align}
	x(k+1) = Ax(k) + Bu(k)
\end{align}
where x represents the temporal activity of 114 brain regions, $A_{114x114}$ is the adjacency matrix quantifying the structural connectivity (see Imaging Data Preprocessing above for details), $u(t)$ represents the input to the system (here the electric charge), and B denotes the brain regions that distribute the input energy across the system (Figure \ref{fig1}c). 

For this system, Modal Controllability (MC) of node $i$ is estimated as $MC_i=\sum_j^{114} [1-\xi_j^2 (A)]v_{ij}^2 $ where $\xi_j$ and $v_ij$ are the eigenvalues and (normalized) eigenvectors of A. Whole-brain modal controllability ($\overline{MC}$) is then defined as the average of nodal controllability over all nodes. Using the symmetry of A this simplifies to: 
\begin{align}
	\overline{MC}=\frac{1}{114}\sum_i^{114}\sum_j^{114}[1-\xi_j^2(A)]v_{ij}^2=
	\frac{1}{114}\sum_j^{114}[1-\xi^2_j(A)]\sum_i^{114}v_{ij}^2=1-\frac{1}{114}\sum_j^{114}\xi^2_j(A)
\end{align}

Note that $\overline{MC}$ is now a function of the eigenvalues only. Thus, given that the eigenvalues determine the decay rate of the response to any arbitrary input $u(t)$, equation (3) implies that whole-brain modal controllability is negatively proportional to the duration of the output signal and therefore to the output power over a fixed interval of time (note that the power of any signal $s(k)$ is defined as $P(u)= \sum_{k=0}^N ||s(k)||^2$  where $N$ is the length of nonzero elements of $s(k)$).

Also, for the system defined in (1), average controllability measures the energy content of the system’s impulse response and is defined as $Tr (W_c)$ where $W_c=\sum_{t=0}^\infty A^t BB^T (A^T)^t $ is the controllability Gramian of the system. Note that for a linear time-invariant system of this form, the impulse response fully determines the system’s response to any (not necessarily impulse) inputs.[13] Average controllability for a single node i (i.e., B includes only one nonzero element in the i’th row) simplifies to:

\begin{align}
	AC_i = Tr(W_c)=Tr\left(B^T\sum^\infty_{t=0}A^{2t}B \right) = Tr\left(B^T\sum_{t=0}^{\infty} V\Xi^{2t}V^TB \right) = Tr(B^TV(I-\Xi^2)^{-1}V^TB) = \sum^{114}_{j=1}\frac{v_{ij}^2}{1-\xi_j^2}
\end{align}
where $V$ is the matrix of eigenvectors and $\Xi$ is a diagonal matrix of eigenvalues. Thus, whole-brain (i.e. mean) average controllability over all nodes becomes
\begin{align}
	\overline{AC}=\frac{1}{114}\sum_{i=1}^{114}\sum_{j=1}^{114}\frac{v^2_{ij}}{1-\xi^2_j} = \frac{1}{114}\sum_{j=1}^{114}\sum_{i=1}^{114}\frac{v^2_{ij}}{1-\xi^2_j} = \frac{1}{114}\sum_{j=1}^{114}\frac{1}{1-\xi_j^2}
\end{align}

Note the similarity between equation (5) and equation (3), where the effects of the eigenvectors disappear when considering the whole-brain (i.e. averaging over all regions) controllability. Also, note the opposite dependence on the system modes where larger eigenvalues increase the whole-brain average controllability and thus, opposite to the whole-brain modal controllability, is associated with longer output response and higher output power. However, we have noticed that the relation between whole-brain average and modal controllability generally follows a parabolic trajectory with extreme values for largest and smallest eigenvalues (data not shown here) and thus the relative nonlinear (nontrivial) difference between whole-brain average and modal controllability is mainly determined by the largest (slowest) and smallest (fastest) modes.

Therefore, following the hypothesis discussed in the introduction, equations (3) and (5) predict that lower whole-brain modal controllability and higher whole-brain average controllability are associated with longer output duration and thus higher signal power of the output (i.e., during the tonic-clonic seizure) and should thus result in higher PSI. Note, however, that the dominating modes in these two metrics are different (slowest dynamics for average and fastest dynamics for modal controllability). Also, the electric charge from ECT (i.e., the control input $u$) has stopped by the time PSI is calculated and only serves (in the model) to bring the brain’s state to a tonic-clonic seizure state from which the decay of the system to the postictal state can be modeled using autonomous (u=0) dynamics.
\newpage

\subsection{Supplementary Material 1: Study inclusion and exclusion criteria}

We recruited Caucasian subjects varying in age from 18-59 years. Patients that were treated predominantly for depressive symptoms were recruited from local psychiatric hospitals. Exclusion criteria comprised any neurological abnormalities, history of seizures, head trauma or unconsciousness, not adequately substituted hypothyroidism, severe physical impairment (e.g. cancer, instable diabetes, epilepsy etc.), pregnancy, claustrophobia, color blindness and general magnetic resonance imaging contradictions (e.g. metallic objects in the body). Further, patients with comorbid life-time diagnoses of schizophrenia, schizoaffective disorder or substance dependence were excluded.

\subsubsection*{Calculation of medication indices}
The Medication Load Index (MedIndex [14]) was calculated as follows:  We defined each psychopharmacological medication as absent (= 0), equal to no medication intake, low (= 1), meaning a dosage equal or lower than average, or high (= 2), with a dosage greater than average relative to the midpoint of the daily dose range recommended by Physician’s-Desk-Reference. We calculated the sum of all medication scores, if patients had more than one prescription. Further, chlorpromazine equivalent doses were calculated for antipsychotic medication load based on Gardner et al. [15].
\newpage
\subsection{Supplementary Material 2:  Description of electroconvulsive therapy}

All patients started their treatment with right-sided unilateral ECT. In eight patients, treatment was converted to bilateral ECT because of insufficient clinical response to unilateral treatment. ECT monitoring included electroencephalogram (EEG), electromyogram (EMG), electrocardiogram (ECG), and blood pressure monitoring. The initial stimulus intensity was calculated using the age method. Re-stimulation, including dosage elevation in steps of 10\%, was considered during single ECT sessions if the primarily induced seizure activity lasted less than 25 seconds in EEG. Due to an increasing seizure threshold throughout the course of ECT, stimulus intensity was increased in the same manner. All patients were anesthetized with methohexital sodium, propofol or sodium thiopental, and a muscle relaxant (succinylcholine) was administered. 
Average Seizure Time was measured by EEG in time wave-seizure activity in seconds. Average Muscle Seizure Time was measured by EMG. During ECT treatment, blood circulation of the left arm was cut off temporarily using a blood pressure cuff before Succinylcholin administration to observe muscle seizure activity. Postictal Suppression Index measures successful inhibition of seizure activity calculated by the ratio of EEG amplitude before and after the seizure ceases. Maximum Stimulus Charge is reported as percentage value of 504 mC (millicoulomb). $\Delta$Stimulus Charge is calculated by subtracting the applied charge during the first ECT from the applied charge during last ECT and is therefore a measure of how much applied charge had to be adjusted, an indicator of poor seizure quality throughout the treatment. Indeces were calculated automatically by the ECT instrument (Thymatron system IV; Somatics Inc).

\begin{table}
	\centering
	\begin{tabular}{l c c c}
		\hline\hline\
		Parameters & N & Mean & SD\\\hline\hline
		Number of ECT treatments & 45 & 13.0 & 4.4\\
		Average Seizure Time (EEG) & 45 & 41.9 & 10.8\\
		Average Seizure Time (EMG) & 45 & 21.1 & 8.5\\
		Postical Suppression Index & 45 & 82.5 & 22.0\\
		Average Stimulus Charge & 45 & 55.7 & 25.4\\
		Maximum Stimulus Charge & 45 & 65.3 & 28.3\\
		Delta Stimulus Charge & 45 & 22.1 & 36.1
	\end{tabular}
\caption{Overview of ECT parameters\\ Note: n = number of included ECT Patients, SD = standard deviation, EEG = electroencephalogram, EMG = electromyography}
\end{table}
\newpage

\subsection{Supplementary Material 3: Medication and comorbid disorder details in patient group}

\begin{center}
	\begin{tabular}{l c}
		\hline\hline\
		Characteristics & ECT$^1 (n=50)$\\\hline\hline
		$T_0$\tiny(before ECT treatment) &  \\\hline
		Medication load & $3.26 \pm 2.08$\\
		Antidepressants & \\
		\quad NaSSA & 8\\
		\quad Tricyclics & 5\\
		\quad NDRI & 1\\
		\quad SSRI & 3\\
		\quad SSNRI & 19\\
		\quad MAO-Inhibitors & 1\\
		\quad Other & 6\\
		Antipsychotics & 23\\
		Mood stabilizer & 7\\\hline
		$T_1$\tiny(after ECT treatment)\\\hline
		Medication load & $3.58\pm 2.29$\\
		Antidepressants & \\
		\quad NaSSA & 10 \\
		\quad Tricyclics & 3\\
		\quad NDRI & 2\\
		\quad SSRI & 2 \\
		\quad SSNRI & 21 \\
		\quad MAO-Inhibitors & 0\\
		\quad Other & 4\\
		Antipsychotics & 22 \\
		Mood stabilizer & 5\\\hline
		$T_0-T_1$ & \\\hline
		$\Delta$ Medication load & $-0.32\pm 1.63$\\
		(repeated measures t-test not significant: t(df=49)=1.36, p=.17)\\
		& \\
		& \\\hline
		Depression subtype & \\\hline
		Psychotic depression & 2\\\hline
		Co-morbid disorders & \\\hline
		GAD & 1\\
		Panic/Agoraphobia & 17\\
		Specific phobia & 2\\
		PTSD & 4\\
		OCD & 1\\
		Social phobia & 7\\
		Dysthymia & 3\\
		Eating disorder & 5\\
		Substance abuse & 12
	\end{tabular}
\end{center}

\newpage
\subsection{Supplementary Material 4}
Measures for outlier detection included 1. average number of streamlines, 2. average fractional anisotropy, 3. average prevalence of each subject’s connections (low value, if the subject has “odd” connections), and 4. average prevalence of each subjects connected brain regions (high value, if the subject misses commonly found connections). For each metric the quartiles (Q1, Q2, Q3) and the interquartile range (IQR=Q3-Q1) was computed across the group and a datapoint was declared as an outlier if its value was below Q1-1.5*IQR or above Q3+1.5*IQR on any of the four metrics. This led to the exclusion of three subjects. Note that results do not substantially change of these three outliers are not removed. 
\newpage
\subsection{Supplementary Material 5}
Code running the PHOTON AI Machine Learning approach applied separately for Fractional Anisotropy, Mean Diffusivity, and Number of Streamlines derived from each patients structural connectome.

\begin{lstlisting}[language=Python,frame=single]
from photonai.base import Hyperpipe, PipelineElement, OutputSettings, Switch
from sklearn.model_selection import LeaveOneOut
	
output_settings = OutputSettings(overwrite_results=True)
pipe = Hyperpipe('photon_ANALYSIS_NAME', 
                 optimizer='grid_search',
                 metrics=['mean_squared_error'],
                 best_config_metric='mean_absolute_error',
                 inner_cv=LeaveOneOut(), 
                 outer_cv=LeaveOneOut(),
                 calculate_metrics_across_folds=True,
                 calculate_metrics_per_fold=False,
                 random_seed=True, verbosity=1,
                 project_folder='./tmp/',
                 output_settings=output_settings)
	
""" add imputer and scaler """
pipe += PipelineElement('VarianceThreshold')
pipe += PipelineElement('SimpleImputer')
pipe += PipelineElement('RobustScaler')
	
""" add transformer elements """
transformer_switch = Switch('TransformerSwitch')
	
transformer_switch += PipelineElement('PCA',
 hyperparameters={'n_components': None}, test_disabled=True)
	 
transformer_switch += PipelineElement('FRegressionSelectPercentile',
 hyperparameters={'percentile': [5, 10]}, test_disabled=True)
	 
pipe += transformer_switch
	
""" add estimator elements """
estimator_switch = Switch('EstimatorSwitch')
estimator_switch += PipelineElement('LinearSVR',
 hyperparameters={'C': [1e-8, 1e-4, 1, 4]})
estimator_switch += PipelineElement('SVR', kernel='rbf')
estimator_switch += PipelineElement('RandomForestRegressor')
estimator_switch += PipelineElement('LinearRegression')
	
pipe += estimator_switch
results = pipe.fit(X, targets)  
\end{lstlisting}
\newpage

\section*{Supplementary References}
1. 	Repple J, Meinert S, Grotegerd D, Kugel H, Redlich R, Dohm K, et al. A voxel-based diffusion tensor imaging study in unipolar and bipolar depression. Bipolar Disord. 2017;19:23–31.

2. 	Lange SC de, Heuvel MP van den. Structural and functional connectivity reconstruction with CATO - A Connectivity Analysis TOolbox. BioRxiv. 2021:2021.05.31.446012.

3. 	Collin G, van den Heuvel MP, Abramovic L, Vreeker A, de Reus MA, van Haren NEM, et al. Brain network analysis reveals affected connectome structure in bipolar I disorder. Hum Brain Mapp. 2016. 2016. https://doi.org/10.1002/hbm.23017.

4. 	Repple J, Mauritz M, Meinert S, de Lange SC, Grotegerd D, Opel N, et al. Severity of current depression and remission status are associated with structural connectome alterations in major depressive disorder. Mol Psychiatry. 2020;25:1550–1558.

5. 	Sarwar T, Ramamohanarao K, Zalesky A. Mapping connectomes with diffusion MRI: deterministic or probabilistic tractography? Magn Reson Med. 2019;81:1368–1384.

6. 	Hagmann P, Cammoun L, Gigandet X, Meuli R, Honey CJ, Van Wedeen J, et al. Mapping the structural core of human cerebral cortex. PLoS Biol. 2008. 2008. https://doi.org/10.1371/journal.pbio.0060159.

7. 	Cammoun L, Gigandet X, Meskaldji D, Thiran JP, Sporns O, Do KQ, et al. Mapping the human connectome at multiple scales with diffusion spectrum MRI. J Neurosci Methods. 2012. 2012. https://doi.org/10.1016/j.jneumeth.2011.09.031.

8. 	Mori S, Van Zijl PCM. Fiber tracking: Principles and strategies - A technical review. NMR Biomed. 2002.

9. 	de Reus MA, van den Heuvel MP. Estimating false positives and negatives in brain networks. Neuroimage. 2013;70:402–409.

10. 	Thomas PJ, Olufsen M, Sepulchre R, Iglesias PA, Ijspeert A, Srinivasan M. Control theory in biology and medicine: Introduction to the special issue. Biol Cybern. 2019;113:1–6.

11. 	Gu S, Pasqualetti F, Cieslak M, Telesford QK, Yu AB, Kahn AE, et al. Controllability of structural brain networks. Nat Commun. 2015;6:1–10.

12. 	Tang E, Bassett DS. Colloquium: Control of dynamics in brain networks. Rev Mod Phys. 2018;90:31003.

13. 	Oppenheim A V, Willsky AS, Nawab SH. Signals and systems Prentice Hall. Inc, Up Saddle River, New Jersey. 1997;7458.

14. 	Redlich R, Almeida JRC, Grotegerd D, Opel N, Kugel H, Heindel W, et al. Brain morphometric biomarkers distinguishing unipolar and bipolar depression: a voxel-based morphometry-pattern classification approach. JAMA Psychiatry. 2014;71:1222–1230.

15. 	Gardner DM, Murphy AL, O’Donnell H, Centorrino F, Baldessarini RJ. International consensus study of antipsychotic dosing. Am J Psychiatry. 2010;167:686–693.

\end{document}